\documentstyle[aps,floats,epsf]{revtex}
\tighten
\draft
\begin{document}

\twocolumn[\hsize\textwidth\columnwidth\hsize\csname
@twocolumnfalse\endcsname

\hfill$\vcenter{\hbox{\bf IUHET-428} \hbox{October 2000}}$  

\title{Anomaly-Free Flavor Symmetry and Neutrino Anarchy} 
 
\author{M.~S. Berger and Kim Siyeon}
\address{Physics Department, Indiana University, Bloomington, Indiana 47405}
\date{October 20, 2000}
\maketitle

\begin{abstract}
We show that one can describe the quark and lepton masses with a single
anomaly-free $U(1)$ flavor symmetry provided a single order one parameter is
enhanced by roughly $4-5$. The flavor symmetry can be seen to arise from 
inside the $E_6$ symmetry group in such a way that it commutes with the
$SU(5)$ grand unified gauge group. The scenario does not distinguish between
the left-handed lepton doublets and hence is a model of neutrino anarchy.
It can therefore account for the large mixing
observed in atmospheric neutrino experiments and predicts that the solar 
neutrino oscillation data is consistent with the large mixing angle solution 
of matter-enhanced oscillations.
\end{abstract}

\pacs{12.15.Ff, 11.30.Hv, 14.60.Pq}
\vspace{0.25cm}]
\narrowtext

\newcommand{\be}{\begin{equation}}
\newcommand{\ee}{\end{equation}}
\newcommand{\bea}{\begin{eqnarray}}
\newcommand{\eea}{\end{eqnarray}}


{\bf Introduction:\ } 
The traditional approach to expressing the CKM matrix and the quark and lepton
matrices is to expand in the small parameter, 
$\lambda \sim |V_{us}|\simeq 0.22$. This is well-justified because the 
experimental data for the mass ratios and CKM elements 
appears to be well-described by integer exponents of this expansion parameter, 
\bea
&&|V_{us}|\sim \lambda \;, \ \ |V_{cb}|\sim \lambda ^2 \;, \ \ 
|V_{ub}| \sim \lambda ^4\;.
\label{ckmelem}
\eea
One also has a constraint on 
the CKM elements from $B_d^0-\overline{B}_d^0$ mixing\cite{pdg},
\bea
&&|V_{tb}^*V_{td}|=0.0084\pm 0.0018\;, \label{pdgdata2}
\eea
which implies that 
\bea
&&|V_{td}|\sim \lambda ^3\;.\label{Vtd}
\label{ckmelem2}
\eea
In the quark sector, one has the mass ratios
\bea
{{m_u}\over {m_c}}\sim\lambda^4,
\ \ \ {{m_c}\over {m_t}}\sim\lambda^4\ \ \ 
{{m_d}\over {m_s}}\sim\lambda^2,
\ \ \ {{m_s}\over {m_b}}\sim\lambda^2\;,  \label{quarks}
\eea
while in the lepton sector, the mass ratios
\bea
{{m_\mu}\over {m_\tau}}\sim\lambda^2,
\ \ \ {{m_e}\over {m_\tau}}\sim\lambda^4\;. \label{chleptons}
\eea
These mass ratios are valid at a high (grand unified) scale and
are consistent with the experimental constraints near the electroweak 
scale after including 
renormalization group scaling\cite{bbo}.
The remaining constraints on leptons involve the neutrino masses and mixings.
The most interesting aspect of the neutrino data is that the atmospheric 
neutrino mixing appears to be large, perhaps even maximal. 
If the neutrino masses also obey some hierarchical relations, it seems 
at first sight to be difficult
to understand such a pattern for the neutrino masses, since large
mixing should result when the neutrino masses are of roughly the same order 
of magnitude. The Super-Kamiokande data\cite{superk} suggest that 
\bea
&&\Delta m^2_{23}\sim 2.2\times 10^{-3}~{\rm eV}^2\;, \quad
\sin ^2 2\theta_{23}^\nu \sim 1\;, \label{atmos}
\eea
where the subscripts indicate the generations of neutrinos involved in the 
mixing (We assume the mixing is between $\nu_\mu$ and $\nu_\tau$, and not
some sterile neutrino. We are also not using the LSND data.).

The solar neutrino flux can be accounted for by distinct solutions.
Two of these involve matter-enhanced oscillation (MSW), while the third
involves vacuum oscillations (VO). The two MSW solutions are differentiated by
the size of the mixing angle, so one is usually called the small mixing angle
(SMA) solution, and the other is called the large mixing angle (LMA) solution.
The most recent data disfavors the VO and SMA solution at the 95\% 
confidence level\cite{superk2,nu2000}.
The rough values required for the mixing parameters in the two MSW cases
are shown in the table below.
\bea
\matrix{&\Delta m_{1x}^2\ [eV^2]&\sin^22\theta_{1x}\nonumber \\ \nonumber
{\rm MSW(SMA)}&5\times10^{-6}&6\times10^{-3}\\ \nonumber
{\rm MSW(LMA)}&2\times10^{-5}&\sim 1}
\eea

Since the most recent data favors the LMA solution for 
solar neutrinos, there has been a great deal of effort recently to explain
the neutrino data with various approaches like bimaximal mixing models or 
neutrino anarchy\cite{anarchy1,ns,anarchy2,sy}. The LMA 
solution is the most interesting from 
the standpoint of neutrino factories\cite{nufact}, but requires us to 
understand how the lepton sector differs in terms of hierarchies (of masses
and mixing angles) from the quark sector. In this note we present a model 
that exhibits neutrino anarchy.

Flavor symmetries have been useful tools for modeling the patterns of 
fermion masses and mixings\cite{flavor}.
The Froggatt-Nielsen mechanism\cite{fn} is a popular method for systematically
generating a hierarchy in the Yukawa couplings.
Conside a horizontal $U(1)_\theta$ symmetry under which the 
Standard Model fields
carry charges. Yukawa interactions are now required to respect this 
horizontal (or flavor)
symmetry, and they can arise in two ways: (1) as renormalizable
interactions $\Psi _i \Psi_j H$, or (2) as nonrenormalizable interactions 
involving a gauge singlet superfield $\theta$ which we can assume without 
loss of generality has a flavor charge $-1$,
\bea
&&\Psi _i \Psi _j H\left ({{\theta }\over {M}}\right )^{n_{ij}}\;.
\eea
For this effective term to respect the flavor symmetry, the charges of the 
fields must sum to zero.
The assumed smallness of the parameter $\epsilon={{<\theta >}\over {M}}$ can 
give rise then to a hierarchy of masses from factors of the form
$\epsilon ^{n_{ij}}$. Following the reasoning given in the previous section,
we take $\epsilon \sim \lambda ^2$ where again 
$\lambda $ is identified with the (sine of) Cabibbo angle, $0.22$.
The Froggatt-Nielsen mechanism does not by itself determine the order one
coefficients. A fundamental theory would presumably fix their values.

A flavor symmetry can suppress the entries in a systematic
way compared to order one entries. 
By assigning charges to the various fields one can obtain Yukawa matrices
in reasonable agreement with the experimentally measured values.

We suggest in this note that the 
true expansion parameter is actually $\epsilon \sim \lambda ^2$, and the 
largeness of $|V_{us}|$ in comparison to $\epsilon$ 
comes about because of a presumably order one
coefficient, $C$, 
that turns out to be of order $4-5$. The same large coefficient
can contribute to $|V_{td}|$ yielding a value $(4-5)\epsilon ^2$ which is 
consistent with previous expansions in terms of $\lambda$, 
Eq.~(\ref{ckmelem2}). This also resolves the problem of the discrepancy between
the relative sizes of $|V_{ub}|$ and $|V_{td}|$ where the former is 
best described by $\lambda ^4$ and the latter is best described by 
$\lambda ^3$; this mismatch has proven to be a challenge to model builders
employing a $U(1)$ flavor symmetry.
In most unified models the quark charges are related
to the  lepton charges and odd exponents appear in the mixing angles 
in the neutrino sector, in disagreement with the data.
It can be overcome but usually requires a more complicated flavor symmetry than
might otherwise be the case. 

If one takes the hierarchy in Eq.~(\ref{ckmelem}) 
seriously and insists on obtaining 
$|V_{us}|\sim \lambda$ and $|V_{td}|\sim \lambda ^3$, 
then there is a unique solution
implementing a $U(1)$ flavor symmetry. This solution is the one found by 
Elwood, Irges, and Ramond\cite{eir}.
The lepton sector can be given charge assignments, and neutrino 
masses and mixings can be derived. We emphasize here that the neutrino 
data prefers even exponents 
of $\lambda$ to best describe the experimental data. The data suggests that 
$\sin \theta _{1x}\sim \lambda ^0$ (which henceforth we will denote simply 
by 1) for the LMA solution, or 
$\sin \theta _{1x}\sim \lambda ^2$ for the SMA solution.
The model gives, however, that 
$\sin \theta_{1x}\sim \lambda ^3$\cite{eir}, a value that is somewhat small
compared to the MSW(SMA) solution. It can be shown that this odd exponent 
results from grand unified relations between quarks and leptons 
and the insistence that there are odd exponents in the CKM 
elements. We show below that the odd exponents
in the quark sector can be seen as arising from just one order one 
coefficient that has fluctuated upward enough to disturb the naive 
hierarchy by one inverse power of $\lambda$.
Furthermore it seems likely that 
if any of the undertermined order one coefficients does
fluctuate to a large value, 
it is most likely to be ones arising in the lighter two generations.

Large mixing for neutrinos is problematic for two reasons: (1) it seems that
without fine tuning, large mixing
should be associated with mass eigenvalues of the same order of magnitude, and
(2) the CHOOZ data\cite{chooz} indicates that one of the three mixing angles
is small ($|U_{e3}| < 0.15$, where $U$ is the neutrino mixing matrix), 
of order $\lambda$ or smaller. Both of these issues have been 
discussed recently in Refs.~\cite{anarchy1,anarchy2} where it was shown that 
it is not so unnatural for a neutrino mass matrix with all entries of order one
to give acceptable mixing angles in agreement with the neutrino data.


{\bf The Anomaly-Free Abelian Flavor Symmetry:\ }
Our approach is to find an anomaly-free flavor symmetry. We assume that this
$U(1)$ symmetry breaks, despite the absence of a Green-Schwarz mechanism, so
as to give a hierarchical mass matrix pattern according to the 
Froggatt-Nielsen mechanism.
The large mixing angles for neutrinos can be achieved by having the 
lepton electroweak 
doublets be indistinguishable under the flavor symmetry. Embedding
these doublets into the ${\bf 5^*}$ multiplet of $SU(5)$ and assigning
flavor charges to the ${\bf 10}$ multiplet, one can then 
assign charges that yield the correct mass ratios for the quarks,
Eq.~(\ref{quarks}).
The charges for the quark fields in $SU(5)$ multilplets are as 
follows ($i=1,2,3$) where we use $\lambda $ as the expansion parameter and
require the flavor charges to be even integers:
\bea
&&{\bf 10}_i\quad (4,2,0)\;, \nonumber \\
&&{\bf 5^*}_i\quad (0,0,0)\;. \label{charges}
\eea
This gives rise to the Yukawa matrices
\bea
&&{\bf U}\sim \pmatrix{ \lambda ^8 & \lambda ^6 & \lambda ^4\cr
                        \lambda ^6 & \lambda ^4 & \lambda ^2\cr
                        \lambda ^4 & \lambda ^2 & 1}
\qquad 
{\bf D}\sim \pmatrix{ \lambda ^4 & \lambda^4 & \lambda ^4\cr
                      \lambda ^2 & \lambda^2 & \lambda ^2\cr
                      1 & 1 & 1}\;, \label{quark}
\eea
for the quark sector. This set of matrices has been suggested 
previously\cite{anarchy2,models}
as giving a reasonable fit to the data apart
from the measured values of  $|V_{us}|$ and $|V_{td}|$ seemed to be enhanced by
roughly one inverse power of $\lambda $.

The enhancement of $|V_{td}|$ and $|V_{us}|$ can easily arise from the same
large order one coefficient.
Given the down-type Yukawa matrix in Eq.~(\ref{quark}), one finds the following
expression for the leading contribution to the CKM matrix elements
\bea
s_{12}^d\sim &&\left ({{d_{12}}\over {\tilde{d}_{22}}}
-{{d_{13}d_{32}}\over 
{\tilde{d}_{22}}}\right )\nonumber \\
&&+{1\over {\tilde{d}_{22}^2}}
\left (d_{11}d_{21}-d_{11}d_{31}(d_{23}+d_{22}d_{32})\right )\nonumber \\
&&-{1\over {\tilde{d}_{22}^2}}(d_{32}d_{12}+d_{13})
\left (d_{21}d_{31}+d_{31}^2(d_{23}+d_{22}d_{32})\right )\;, \label{mixang}
\eea
where $d_{ij}={\bf D}_{ij}/{\bf D}_{33}$ and 
$\tilde{d}_{22}=d_{22}-d_{23}d_{32}$.
In Eq.~(\ref{mixang}), $s_{12}^d$ is expressed for the case in which all 
of the mixing angles in the right-handed transformation matrix needed to 
diagonalize ${\bf D}$ are large as in Eq.~(\ref{quark}).
This generalized expressions given in the literature\cite{mixing,bk} for which 
only the mixing angle in the second and third generation is of order one.
The CKM matrix can be characterized by four mixing angles which 
are given to leading order by
\bea
&&|V_{us}|= s_{12}^d-s_{12}^u\;, \nonumber \\
&&|V_{cb}|=s_{23}^d-s_{23}^u\;, \nonumber \\
&&|V_{ub}|=s_{13}^d-s_{13}^u-s_{12}^u|V_{cb}|\;, \nonumber \\
&&|V_{td}|=-s_{13}^d+s_{13}^u+s_{12}^d|V_{cb}|\;.
\eea
There are phases in the Yukawa entries which do not affect the expansion 
in terms of $\lambda$, and so we omit them here.
A large order one coefficient that enhances $s_{12}^d$ from its expected
magnitude of $\lambda ^2$ to order $\lambda $ will enhance just those 
CKM elements that the data tells us are large.
One can easily see that the large order one coefficient $C$ contributes to 
$|V_{us}|$ and $|V_{td}|$ by looking at the constraints from the 
unitarity of the CKM matrix. Consider the following unitarity relation
\bea
V_{ud}V_{ub}^*+V_{cd}V_{cb}^*+V_{td}V_{tb}^*&=&0\;.\label{unit}
\eea
From this relation one immediately sees that an enhancement of $|V_{cb}|$ by 
a factor $4-5$ that doesn't at the same time enhance 
$|V_{ub}|$ must be compensated by an enhancement of $|V_{td}|$.

\begin{center}
\vspace*{-1.2in}
\epsfxsize=2.8in
\hspace*{-0.5in}
\epsffile{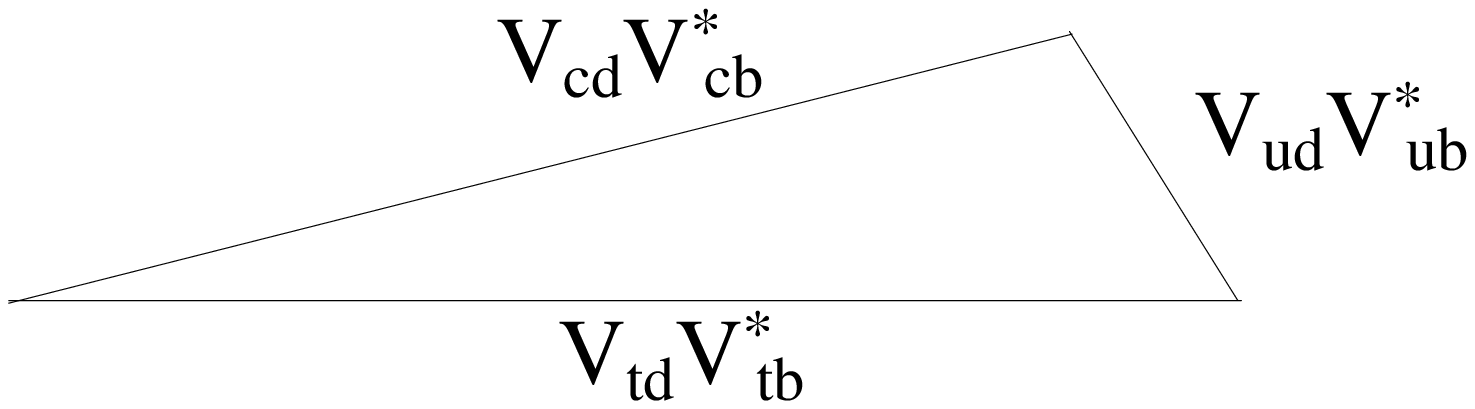}

\vspace*{-0.1in}
\parbox{3in}{\small  Figure 1: Unitarity triangle corresponding to 
Eq.~(\ref{unit}).}
\end{center}

The Yukawa matrix for the charged lepton sector can be deduced from the 
charge assignments in 
Eq.~(\ref{charges}),
\bea
&&{\bf E}\sim \pmatrix{ \lambda ^4 & \lambda ^2 & 1\cr
                        \lambda ^4 & \lambda ^2 & 1\cr
                        \lambda ^4 & \lambda ^2 & 1}\;,
\eea
which yields good agreement with the known mass ratios.
The indistinguishability of the ${\bf 5^*}$ multiplets with respect to the
$U(1)$ flavor symmetry gives rise to large mixing in the lepton sector.


{\bf Embedding in {\boldmath $E_6$}:\ }
In the usual chain of embedding $SO(10)$ inside $E_6$ and $SU(5)$ inside
$S0(10)$, there arise two $U(1)$ symmetries\cite{hr}:
\bea
&&E_6\to SO(10)\times U(1)_\psi\;, \nonumber \\
&&SO(10)\to SU(5)\times U(1)_\chi\;.
\eea
Both of these $U(1)$ symmetries as well as any linear combination is 
guaranteed to be anomaly free by virtue of that fact that $E_6$ is an 
anomaly free group.
We can assign charges to the fields in a way 
that is $SU(5)$-symmetric, and that 
also respects the $U(1)$ flavor symmetry.

The charges of the SU(5) multiplets under the $U(1)_\psi$ and $U(1)_\chi$
are 
\bea
&&\begin{array}{c@{\quad}c@{\quad}c@{\quad}c@{\quad}c@{\quad}c@{\quad}c
@{\quad}c@{\quad}c}
 & {\bf 10_a} & {\bf 5^*_a} & {\bf 1_a} & & {\bf 5_b} & {\bf 5^*_b} & & 
{\bf 1_c} \\
2\sqrt{10}Q_\chi & -1 & 3 & -5 & & 2 & -2 & & 0 \\
2\sqrt{6}Q_\psi & 1 & 1 & 1 & & -2 & -2 & & 4 
\end{array}\;.
\eea
The charges are grouped into the ${\bf 16}$, ${\bf 10}$ and ${\bf 1}$ 
representations of $SO(10)$ inside a ${\bf 27}$ representation of $E_6$.
Now it is easy to see that the charge assignments given in the previous section
are obtained by taking the linear combination
$Q_\theta=3\times(2\sqrt{6}Q_\psi)/16 - (2\sqrt{10}Q_\chi)/16$.
This yields the $U(1)_\theta$ charges
\bea
&&\begin{array}{c@{\quad}c@{\quad}c@{\quad}c@{\quad}c@{\quad}c@{\quad}c
@{\quad}c@{\quad}c}
 & {\bf 10_a} & {\bf 5^*_a} & {\bf 1_a} & & {\bf 5_b} & {\bf 5^*_b} & & 
{\bf 1_c} \\
4Q_\theta & 1 & 0 & 2 & & -2 & -1 & & 3
\end{array}\;. \label{assign} 
\eea
The charge assignments in Eq.~(\ref{charges}) can now be identified as
$(16Q_\theta,8Q_\theta,0)$. This does not then 
represent a strict embedding of the 
representations inside $E_6$, but since each generation has $U(1)_\theta$
charges that are multiples (modings) 
of the $SU(5)\times U(1)_\theta$ inside $E_6$, the
gauge anomalies are guaranteed to be absent as long as the extra matter 
content of the ${\bf 27}$ is present (either near the electroweak scale or
at some higher scale). The construction of the flavor charge from the two
$U(1)$ subgroups of $E_6$ guarantees that the couplings 
${\bf 10_a-5^*_a-5^*_b}$ and ${\bf 5_b-5^*_b-1_c}$ are allowed by the 
$U(1)_\theta$ flavor symmetry.

As is well-known, supersymmetric
gauge coupling unification requires that in addition 
to complete multiplets of $SU(5)$ with masses
at the electroweak scale, there must be 
a pair of states with the quantum numbers of the 
two Higgs doublets of the Minimal Supersymmetric Model (MSSM). These should
arise outside of three generations of ${\bf 27}$ giving rise to the 
Standard Model matter multiplets and the exotic ${\bf 5_b}$, ${\bf 5^*_b}$ and 
${\bf 1_c}$. These Higgs fields could arise\cite{rizzo}
as components from either a 
${\bf 27+27^*}$ or a ${\bf 78}$. A particularly interesting possibility
for the $U(1)_\theta$ charges occurs in the former case where one can choose 
these Higgs fields from the ${\bf 5^*_a} ({\bf 5_a})$ of $SU(5)$ inside the 
${\bf 16} ({\bf 16^*})$ of $SO(10)$. Then the (unnormalized) 
$Q_\chi$ and $Q_\psi$ charges
are ${\bf 5^*_a}(3,1)$ and ${\bf 5_a}(-3,-1)$, so that the Higgs fields have 
vanishing $Q_\theta$ charge (like the lepton doublets in Eq.~(\ref{assign})).

From the charge assignment for the $SU(5)$ singlet state inside the $\bf 16$ 
of $SO(10)$, one obtains the Majorana mass matrix of the right handed 
neutrinos
\bea
{\bf M_R}\sim \pmatrix{ \lambda ^{16} & \lambda ^{12} & \lambda ^8\cr
                        \lambda ^{12} & \lambda ^8 & \lambda ^4\cr
                        \lambda ^8 & \lambda ^4 & 1}M_P\;,
\eea
and the Yukawa matrix for the Dirac masses of the neutrinos
\bea
&&{\bf Y_\nu}\sim \pmatrix{ \lambda ^8 & \lambda ^4 & 1\cr
                          \lambda ^8 & \lambda ^4 & 1\cr
                          \lambda ^8 & \lambda ^4 & 1}\;.
\eea
(This strong hierarchy in the Majorana mass matrix is useful in explaining
the baryon asymmetry of the universe via the process of 
leptogenesis\cite{lepto}.)
Finally via the seesaw mechanism 
${\bf m_\nu}={\bf Y_\nu}^T{\bf M_R}^{-1}{\bf Y_\nu}v_2^2$, where $v_2$ is 
the vacuum expectation value of the Higgs field that couples to
the up-type quarks,
one obtains the mass matrix of the light neutrinos
\bea
{\bf m_\nu}\sim \pmatrix{ 1 & 1 & 1\cr
                        1 & 1 & 1\cr
                        1 & 1 & 1}{{v_2^2}\over {M_P}}\;.
\eea
Of course, it is not necessary that the right handed neutrinos have the 
charge assignment in Eq.~(\ref{assign}) to obtain this light neutrino mass
matrix. It is determined solely by the universal flavor charge assigned to 
the ${\bf 5^*_a}$ multiplets in Eq.~(\ref{charges}).


{\bf Neutrino Oscillations:\ } Large mixing in the atmospheric neutrino
data and LMA solution for solar neutrinos together with the constraint 
from CHOOZ indicates that the neutrino mixing matrix is of the form
\bea 
&&U\sim \pmatrix{1 & 1 & \lambda ^k  \cr
                 1 & 1 & 1  \cr
                 \lambda ^k & 1 & 1 }\;. \label{nmix}
\eea
when expanded in powers of $\lambda $. 
However the naive expectation from power counting for the $U(1)_\theta$ \
symmetry indicates that the 
exponent is $k=0$.
The model seems to suggest that all three mixing angles are large, and 
the largeness of $U_{e3}$ element is problematic in light of the 
CHOOZ reactor data\cite{chooz}, which places an upper limit on the mixing 
$U_{e3}$ from the
constraints on $\overline{\nu}_e$ disappearance.
It appears that, in terms
of the expansion in terms of $\lambda $, this element of the mixing matrix
in Eq.~(\ref{nmix}) should be of order $\lambda $ or smaller to account for 
the lack of $\overline{\nu}_e$ disappearance in the reactor neutrino 
experiment.

Additionally it seems that such a neutrino mixing matrix implies large
mixing between all three generations and neutrinos with masses which do not
exhibit a hierarchy. One approach to remedy this situation is to introduce 
an additional discrete component to the flavor symmetry which can enhance 
or suppress mass eigenvalues or mixing angles in comparison to their values
in a model in which the flavor symmetry is simply $U(1)$\cite{bk,gns}.
However it has been argued recently\cite{anarchy1,anarchy2} 
that it is not so improbable that experimentally acceptable values for 
neutrino masses and mixings can result, even when naively on the basis of 
power counting the neutrino mixing matrix appears to give $U_{e3}$ typically
of order one.
This scenario has been dubbed neutrino anarchy.
It is argued in Ref.~\cite{anarchy2}, where a proper weighting of the 
random order one coefficients has been justified on the assumption of 
basis-independence of the neutrino states, one can, without much fine tuning,
find a result where $U_{e3}$ is just below the current experimental limit.
Indeed, this can be considered the most characteristic feature of neutrino
anarchy that can be experimentally tested in the near future.
While an acceptable value for $U_{e3}$ occurs only in a 10\% tail of its
probability distribution\cite{anarchy2}, the other observables do not need 
any fine-tuning, and the argument that has developed is that this situation
is not an improbable one.


{\bf Conclusions:\ } We have suggested in this note 
that if one is willing to give up the 
assumption that $|V_{us}|\sim \lambda$ is the correct expansion parameter
for the hierarchies evidenced in the fermion masses and mixings, then one 
can get a complete description of the masses and mixings
in the quark, charged lepton, and
neutrino sectors. A particular model in which the $U(1)$ horizontal symmetry
is nonanomalous is easily constructed by taking the appropriate linear 
combination of the $U(1)_\psi$ and $U(1)_\chi$ subgroups of $E_6$.
The CKM elements satisfy $|V_{us}|\simeq C\lambda^2$ and
$|V_{ts}|\simeq C\lambda ^4$ where $C$ is a relatively large order one 
parameter of $4-5$, and the neutrino sector that results falls into the 
class exhibiting the neutrino anarchy of Refs.~\cite{anarchy1,anarchy2}.

More generally we feel an interpretation of the hierarchy in terms of a 
expansion parameter $\epsilon \sim \lambda^2$ is quite reasonable and makes
the resulting model-building much easier since only even powers of $\lambda$ 
will appear. 

The set of $U(1)$ flavor charges for the Standard Model particles has 
appeared in the literature. Here we have shown how to embed this $U(1)$ 
flavor symmetry as a subgroup of the $E_6$ symmetry that commutes with
the standard $SU(5)$ subgroup. Together with recent insights on the 
viability of models where the three lepton doublets are indistinguishable,
the model is in reasonable agreement with all the current experimental data.
 

\vspace{0.5cm}


This work was supported in part by the U.S.
Department of Energy
under Grant No. 
No.~DE-FG02-91ER40661.



\end{document}